\documentclass[prc,superscriptaddress,showpacs,twocolumn,amssymb,amsmath,amsfonts,aps]{revtex4}

\usepackage{graphicx} % Include figure files
\usepackage{dcolumn}  % Align table columns on decimal point
\usepackage{longtable}% Allow column and page-spanning tables.
\usepackage{bm}       % bold math

\begin{document}

%\preprint{CMU/JLab/100-2010}

%\input{authors}

\title{Scaling and Resonances in Elementary $K^+ \Lambda$ Photoproduction}

\author{R. A. Schumacher}
\affiliation{Department of Physics, Carnegie Mellon University, Pittsburgh, PA 15213}
\author{M. M. Sargsian}
\affiliation{Department of Physics, Florida International University, Miami, FL 33199}
 
\date{\today}  % It is always \today

%DRAFT - NOT FOR DISTRIBUTION - DRAFT

\begin{abstract} 
Recent cross section data for the reaction $\gamma + p \to K^+ +
\Lambda$ are examined for evidence of scaling in both the low-$t$
Regge domain and in the high $\sqrt{s}$ and $-t$ domain where
constituent counting may apply.  It is shown that the reaction does
scale in both regimes. At large center of mass angles, $s^{-7}$ scaling
appears to hold at essentially all $-t$, but with angle-dependent
oscillations.  The scaled data show particularly strong evidence for
$s$-channel resonances for $-t$ below 2 GeV$^2$ and for $W$ below
about 2.3 GeV.  The dominant contributions are consistent with an $N^*$
$S_{11}$ resonance at 1690 MeV, a $P_{13}$ at 1920 MeV, and a
$D_{13}$ resonance at 2100 MeV, which interfere to give the observed
strong angular dependence.

\end{abstract}

 % end of PACS codes
%\maketitle
%\pacs{
%      {25.20.Lj}{ Photoproduction reactions}
%      {13.40.-f}{ Electromagnetic processes and properties},
%      {13.60.Le}{ Meson production},
%      {14.20.Gk}{ Baryon resonances with S=0},
%      {13.30.-a}{ Decays of baryons},
%      {13.30.Eg}{ Hadronic decays},
%      {13.60.-r}{ Photon and charged lepton interactions with hadrons},
%     } % end of PACS codes
\maketitle

\section{\label{intro}INTRODUCTION}

In the GeV energy domain, there are competing pictures for how meson
photoproduction reactions can be most economically described.  In the
meson-baryon picture, pseudo-scalar meson photoproduction proceeds by
$s-$ and $u-$channel baryon exchange, plus $t-$channel meson exchange.
At $|-t| < 1$ GeV$^2$ and high $W (=\sqrt{s})$, cross sections are
frequently well described in an approach with the exchange of one or
more Regge trajectories, corresponding to, in this reaction, $K$ and
$K^*$ trajectories~\cite{lag1,lag2}.  For a single trajectory, the
cross section $d\sigma/dt$ can be parametrized
\begin{equation}
d\sigma/dt = D(t)(\frac{s}{s_0})^{2\alpha(t)-2},
\label{eq:1}
\end{equation}
where $s_0$ is a scale factor taken to be 1 GeV and $D(t)$ is a
function of $t$ alone.  $\alpha(t)$ is the Regge trajectory that
describes how the angular momentum of the exchange varies with $t$.
Note that in a reaction where $\alpha(t)\sim 0$, one would expect the
cross section to ``scale'' with $s^{-2}$.  That is, the $s^2$-scaled
cross section would exhibit a uniform smooth dependence on $-t$ that
depends neither on $s$ nor separately on the production angle of the
meson.

Exclusive scattering in the high momentum and energy transfer limit is
thought to follow the constituent counting rules introduced in
Refs.~\cite{brodsky,MMT}.  While the existence of these rules can be
proved rigorously within perturbative QCD (pQCD)~\cite{brodsky}, they
can also be derived based on more general grounds of the constituent
nature of scattering without requiring the validity of
pQCD~\cite{ILS}.  In both interpretations, however, the onset of
constituent counting manifests the transition from peripheral Regge
type scattering to short range hard scattering involving a minimal
number of partonic constituents, plus leptonic or photon fields, via
which a given exclusive reaction can occur.  The constituent counting
rule predicts the analytical form for the differential cross section
$d\sigma/dt$ to be
\begin{equation}
{d\sigma/ dt} = f({t/s}) s^{2-n}
\label{eq:2}
\end{equation}
in the limit that $s\gg M_i^2$ and $t/s$ is fixed, where the $M_i$'s
are the masses of the particles involved in the reaction.  The power
factor $n$ is the minimal number of point-like constituents needed to
accomplish the reaction.  For photoproduction of pseudoscalar mesons,
the relevant $n$ is 9 if the photon is counted as a single elementary
field, so the expectation is that the cross section should scale as
$s^{-7}$.  The requirement that $t/s$ remain fixed at given high $s$
amounts to the meson production angle, $\cos \theta_{c.m.}$, being
held fixed.  The form of the function $f$ is not specified, but can be
in principle calculated either within pQCD or in non-perturbative
models of constituent quark scattering.  The major problem with pQCD,
however, is in significant underestimation of the absolute magnitude
of the function $f$ (see {\it e.g.} Ref.~\cite{Farrar:1990eg}).  Note
that in several instances the energy dependence of hard exclusive
reactions can be reproduced within phenomenological models invoking
only hadronic degrees of freedom ({\it e.g.} in
Ref.~\cite{Laget:2009hs}).  In such models, however, the power law of
the energy dependence is rather an accidental result.  This situation
can not explain the ``persistent'' consistency of the constituent
scaling law observed for many hard exclusive reactions including
hadronic~\cite{White:1994tj} and
photoproduction~\cite{Anderson:1976ph} reactions, involving proton,
deuteron and even $^3He$ targets at $|t|,|u| > M_i^2$ kinematics (see
{\it e.g.} Refs.~\cite{Bochna,Mirazita,Pomerantz}]).  In the
photoproduction case, $s^{-7}$ scaling was found to be consistent with
data for the final states $\pi^+ n$, $\pi^0 p$, $\pi^-\Delta^{++}$,
$\rho^0 p$, and (with poor statistics) for $K^+\Lambda$ and
$K^+\Sigma^0$.  The data in the present study are at $W$ values
covering the $N^*$ resonance domain, but with higher statistics than
earlier work.

At kinematics similar to the reaction studied in the present paper,
pion photoproduction at large angles exhibits $s^{-7}$ scaling when
the transverse momentum in the c.m. frame exceeds 1.2 GeV/c, as well
as possible ``oscillatory'' features around the scaling
prediction~\cite{zhu,chen}.  The former behavior has been interpreted
as a clear signature for the onset of constituent scaling.  The latter
behavior has been discussed in terms of the breakdown of locality in
quark-hadron duality that relates resonance excitations at low
energies to parton phenomena at high energies~\cite{zhaoclose}.  The
analogous $KY$ behaviors have not previously been examined.

In the lower-energy domain of the nucleon resonances, that is, below
$W=\sqrt{s} \sim 2.5$ GeV, several non-strange $I=0$ baryon resonances
contribute to the $\gamma +p \to K^+ + \Lambda$ reaction mechanism.
So-called hadrodynamic models based on effective Lagrangians have, for
many years, been employed with moderate success to describe a wide
range of hadronic and electromagnetic reactions, including the
particular reaction that is the focus of this paper~\cite{sarantsev,
anisovich, anisovich2,Nikonov:2007br,Anisovich:2010an, jan,
jan_a,ireland, corthals, saghai1, mart, maid}.  Looking for such
resonances in strangeness-containing final states has been a hunting
ground for so-called ``missing'' resonances predicted in quark models
~\cite{Capstick:1998uh} but not seen in pionic final state
experiments.  Separating out the resonant contributions to the overall
reaction mechanism has been pursued in various model approaches.  In
recent times, the most advanced methods include unitary
coupled-channels methods that fit data sets from multiple channels
simultaneously. For example, in the approach used by the Bonn-Gatchina
group~\cite{Nikonov:2007br,Anisovich:2010an}, the dominant partial
waves in the present reaction of interest are consistent with
$N(1720)P_{13}, N(1900)P_{13}, N(1840)P_{11}$, plus an $S_{11}$ wave.
A problem with this and similar hadrodynamic approaches lies in the
broad freedom in the overlap of several contributing $N^*$ resonances
and insufficient experimental constraints from spin observables to
uniquely describe the reactions.  As an alternative to the $N^*$
resonance picture involving 3-quark resonances, it has also been
proposed~\cite{MartinezTorres:2009cw,Jido:2009gp} that the
intermediate excitation in this reaction is a $K\overline{K}N$
structure that is dynamically generated in the rescattering of
distinct mesons and baryons.

To address some of these issues, the CLAS Collaboration has published
high-statistics cross section data for the reaction $\gamma + p \to
K^+ + \Lambda$ in recent years~\cite{bradforddsdo,McCracken:2009ra}.
The earlier paper by Bradford {\it et al.}~\cite{bradforddsdo} showed
that the cross section scales at low $-t$ with $s^{-2}$, consistent
with the idea the $\alpha(t)\sim 0$ in this kinematic domain.  Results
in the more recent paper by McCracken {\it et
al.}~\cite{McCracken:2009ra} are entirely consistent with the earlier
paper, but extend the range of $W$ by about 300 MeV, to 2.8 GeV, and
has improved statistical precision by a factor of about four.  This
allows us to revisit the question of scaling and resonances outlined
above.

The paper is organized first with more theoretical background about
scaling in Section~\ref{sec:partonic}, and then the experimental
results demonstrating scaling behavior are shown in
Section~\ref{sec:scaling}.  The apparent resonant aspects of the
scaled data are presented in Section~\ref{sec:resonance} together with
a model description.  The results are discussed and summarized in
Section~\ref{sec:conclusion}.

\section{\label{sec:partonic}PARTONIC CONTENT OF THE REACTION}

The onset of the energy scaling of the differential cross section of
the $\gamma N \rightarrow M B$ photoproduction reaction in the form of
$s^{-N}$ at fixed $\cos \theta_{c.m.}$ indicates in a certain degree the
factorization of the hard sub-processes in the scattering amplitude
which can be expressed in the form (see {\it e.g.}
Refs.~\cite{brodsky, SBB, KSPS, GS})
\begin{eqnarray}
\label{eq:ampl}
M^\lambda(s,t) = \int d^4[k_2] d[p_2] d^4[k_1] d[p_1]
\psi_{M}^\dagger(k_2)\psi_{B}^\dagger(p_2) \\
\times H^\lambda(p_2,k_2,p_1,k_1) 
\psi_{\gamma,phys}(k_1)\psi_{N}(p_1), \nonumber
\end{eqnarray}
where $\psi_{M}$, $\psi_{B}$, $\psi_{N}$ and $\psi_{\gamma,phys}$
represent the soft partonic wave functions of the meson, produced
baryon, initial nucleon and physical photon, respectively.  The
kernel, $H$ represents the amplitude of the factorized hard scattering
sub-process which at asymptotic energies defines the {\em whole energy
dependence} of the total scattering amplitude in the scaling form of
$s^{-\frac{n-4}{2}}$, where $n=n_N+n_{\gamma,phys} + n_{M} + n_{B}$
with $n_{N}$, $n_{\gamma,phys}$, $n_{B}$, $n_{M}$ being the number of
the partons (or elementary fields) entering in the wave function of
the scattering particles.  Note that at pre-asymptotic energies the
above energy dependence is convoluted with the energy dependence
following from the sub-leading as well as non-perturbative
(resonating) processes. The latter can result to the additional
oscillatory energy dependence of the $s^N$ scaled differential cross
sections (see {\it e.g.} Refs.~\cite{BT,RP}).
   
While the number of the partonic constituents in the wave function of
hadrons in Eq.~(\ref{eq:ampl}) can be identified with that of the valence
quarks, the interpretation of the physical photon at pre-asymptotic
energies requires the consideration of both bare and hadronic
components of its wave function.  It is rather well established that
the physical photon's wave function can be represented through the
superposition of a bare photon and hadronic components (see {\it e.g.}
Ref.~\cite{BSYP}):
\begin{equation}
\psi_{\gamma,phys} = \psi_{\gamma} + \psi_{hadron},
% \label{photon}
\end{equation}
where the hadronic part is dominated by intermediate vector mesons
states.  Due to the large interaction cross section the hadronic part
of the photon wave function dominates in many photoproduction
processes especially involving the production of vector mesons.

In the high energy and momentum transfer limit one expects that the
bare photon component will gradually dominate in the photoproduction
cross section since the hard kernel, $H$, involving the hadronic
component of the photon is suppressed by an additional factor,
$s^{-\frac{1}{2}}$, as compared to the amplitude involving the bare
photon only.  However for the intermediate range of energies the
hadronic component may still dominate in the hard processes at
$|t|,|u| >  M_N^2$ due to the relatively large coupling constant of
the physical photon to the vector mesons.  In this case the hard
rescattering (see {\it e.g.} Ref.~\cite{gdpn}) of intermediate vector
mesons off the target nucleon defines the energy dependence of the
photodisintegration cross section.

This situation can explain the observed $s^{-8}$ scaling of hard real
Compton scattering at large $\cos \theta_{c.m.}$~\cite{hcomp} as well as
exclusive photoproduction of $(\rho+\omega)$
mesons~\cite{Anderson:1976ph} which agrees reasonably well with the
$s^{-8}$ scaling of the differential cross section.

The situation however is simplified with consideration of the
photoproduction of pseudoscalar mesons. In this case the contribution
from $\psi_{hadron}$ is suppressed since the dominating $vector \
meson + N\rightarrow pseudoscalar \ meson + N^\prime$ rescattering
will proceed through the double helicity flip scattering which one
expects to be suppressed in the high momentum transfer limit.  This
expectation is confirmed in the exclusive photoproduction of
$\pi$-mesons in $\gamma N\rightarrow \pi N$ reactions which clearly
shows a $s^{-7}$ scaling starting already at $s \approx 7-8$~GeV$^2$
\cite{Anderson:1976ph,chen,zhu}.

Similar (early) scaling, may be expected for kaon photoproduction in
$\gamma + p\rightarrow K^+ + \Lambda$ or $\gamma + p\rightarrow K^+ +
\Sigma^0$ reactions.  The precision of the previous
data~\cite{Anderson:1976ph} on exclusive kaon photoproduction was not
sufficient ($s^{-N}$, $N = 7.3\pm 0.4$) to rule out unambiguously the
contribution from the hadronic component of the photon wave function,
which results in $N=8$.

\section{\label{sec:scaling}SCALING OF THE CROSS SECTION}

\begin{figure*}
\resizebox{1.0\textwidth}{!}{\includegraphics[angle=0.0]{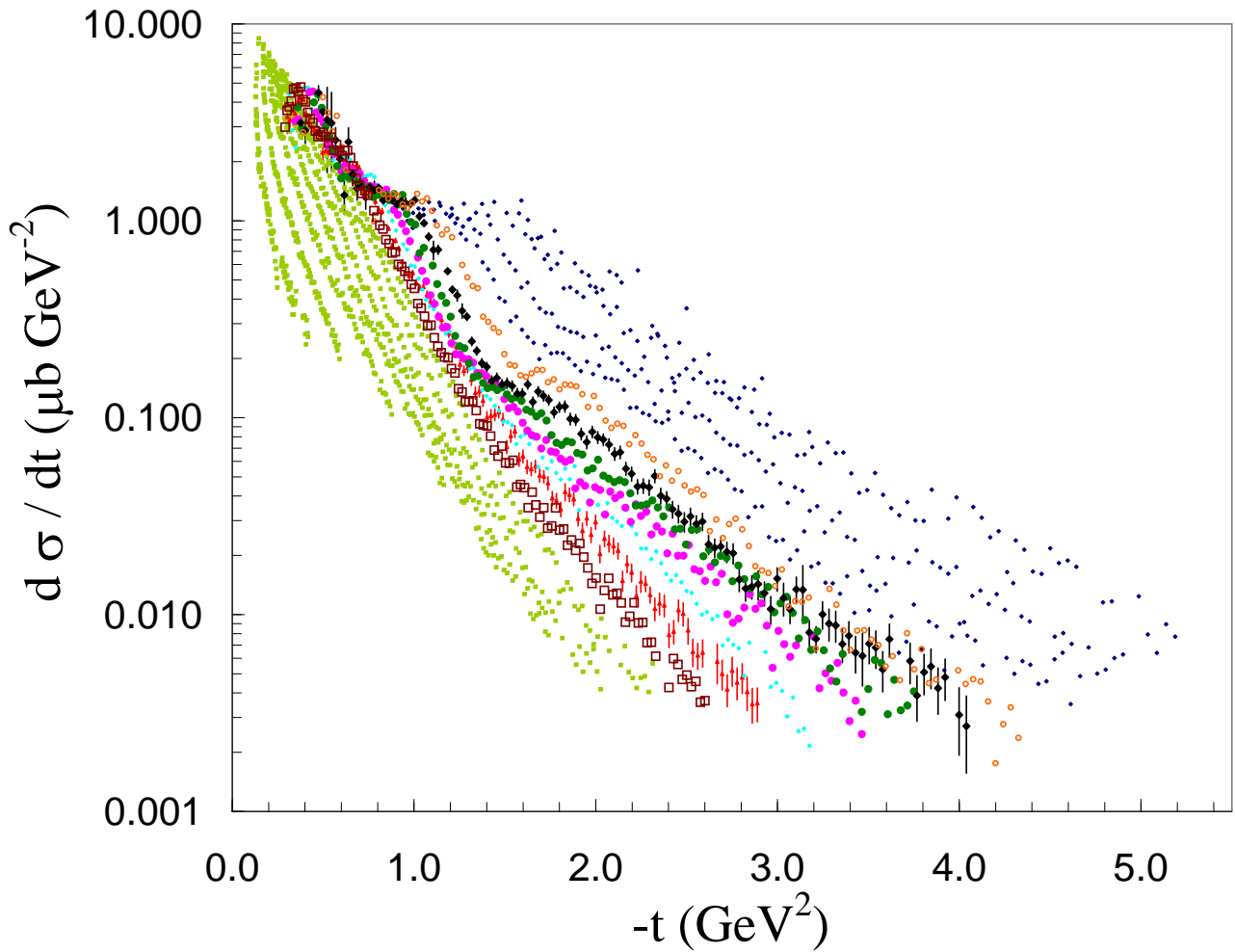}}
\caption{ (Color online) Cross section for the reaction $\gamma + p
\to K^+ + \Lambda$ as a function of $-t$ with no scaling factors
applied. Each band of points shows data for a bin in
$\Delta\cos\theta_{c.m.}$ of 0.1.  Only a few specific bands have been
highlighted with different colors.  The color code is: all forward
angle bands for $\cos\theta_K$ $+0.9$ to $+0.2$ (light green points),
$+0.1$ (open brown squares), 0.0 (filled red triangles), $-0.1$
(filled cyan points), $-0.2$ (filled magenta circles), $-0.3$ (filled
green squares), $-0.4$ (filled black diamonds), $-0.5$ (open orange
circles), $-0.6$ to $-0.9$ (blue points).  Representative statistical
error bars are shown for only a few angle bands for clarity.  }
\label{fig:dsdt}       % Give a unique label
%\end{figure}
\end{figure*}

Figure~\ref{fig:dsdt} shows the complete set of differential cross
sections $d\sigma/dt$ from Ref.~\cite{McCracken:2009ra} versus $-t$.
The published data were transformed from $d\sigma/d\cos\theta_{c.m.}$
in kaon production angle to $d\sigma/dt$ with the Jacobian $1/2kq$,
where $k$ is the initial state c.m. momentum and $q$ is the final
state c.m. momentum.  The data are binned in bands of center of mass
angle $\cos\theta_{c.m.}$, each of width 0.1.  Representative error
bars are shown only for several angles important for this discussion;
the close spacing of the data points makes it easy to trace the trends
as a function of angle.  The important observation here is that for
forward angles (green points) the cross sections fall smoothly with
increasing $-t$, and that there is some hint of ``structure'' for
intermediate angles near $-t\sim 1.0$ GeV$^2$.

%%%%%%%%%%%%%%%%%%%%%%%%%%Fig 1%%%%%%%%%%%%%%%%%%%%%%%%%%%%%%%%%%%%
%remove the ``*'' to put the figures in-line with text.  
%\begin{figure}
%\resizebox{0.50\textwidth}{!}{\includegraphics[angle=0.0]{Scaling_FigA.pdf}}
%%%%%%%%%%%%%%%%%%%%%%%%%%Fig 1%%%%%%%%%%%%%%%%%%%%%%%%%%%%%%%%%%%%

\begin{figure*}
\resizebox{1.0\textwidth}{!}{\includegraphics[angle=0.0]{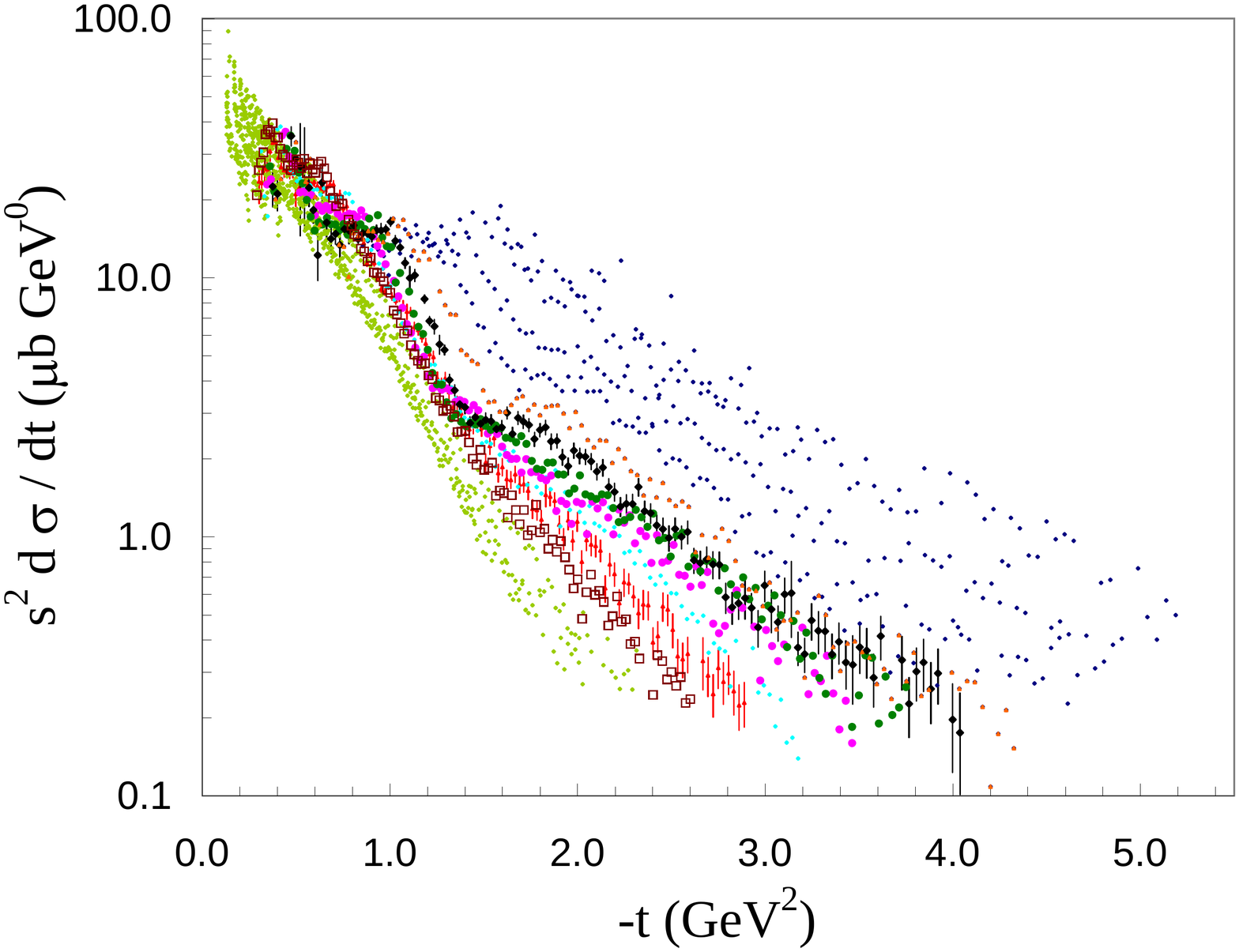}}
\caption{ (Color online) Cross section for the reaction $\gamma + p
\to K^+ + \Lambda$ as a function of $-t$ with a scaling factor of
$s^2$ applied.  Note how the forward angle points (light green) fall
approximately on a single locus.  The data and color coding are the
same as for Fig.\ref{fig:dsdt}.  }
\label{fig:dsdts2}       % Give a unique label
%\end{figure}
\end{figure*}

Figure~\ref{fig:dsdts2} shows the same data with a scaling factor of
$s^2$ applied.  It is evident that the forward-angle data now fall on
a fairly tight locus of points, while for $|-t| \ge 1.0$ GeV$^2$ this
simple scaling fails.  The scaling exponent of 2 is qualitatively
optimal.  The function $D(t)\sim e^{bt}$ in Eq.~\ref{eq:1} has a slope
estimated as $b=3.0\pm 0.7$ GeV$^{-2}$.  Refining the effective Regge
trajectory as per $\alpha(t) = \alpha_0 + \alpha^\prime t$, and
adjusting the value of $-t$ at which the trajectory ``saturates'',
results in a somewhat tighter bunching of the loci of points, but for
the present discussion we chose not to fine-tune this approach.  This
``Regge-scaling'' of the low $-t$ data simply confirms the
observations made in Ref.~\cite{bradforddsdo}.

%%%%%%%%%%%%%%%%%%%%%%%%%%Fig 2%%%%%%%%%%%%%%%%%%%%%%%%%%%%%%%%%%%%
%remove the ``*'' to put the figures in-line with text.  
%\begin{figure}
%\resizebox{0.50\textwidth}{!}{\includegraphics[angle=0.0]{Scaling_FigB.pdf}}
%%%%%%%%%%%%%%%%%%%%%%%%%%Fig 2%%%%%%%%%%%%%%%%%%%%%%%%%%%%%%%%%%%%

\begin{figure}
\resizebox{0.50\textwidth}{!}{\includegraphics[angle=0.0]{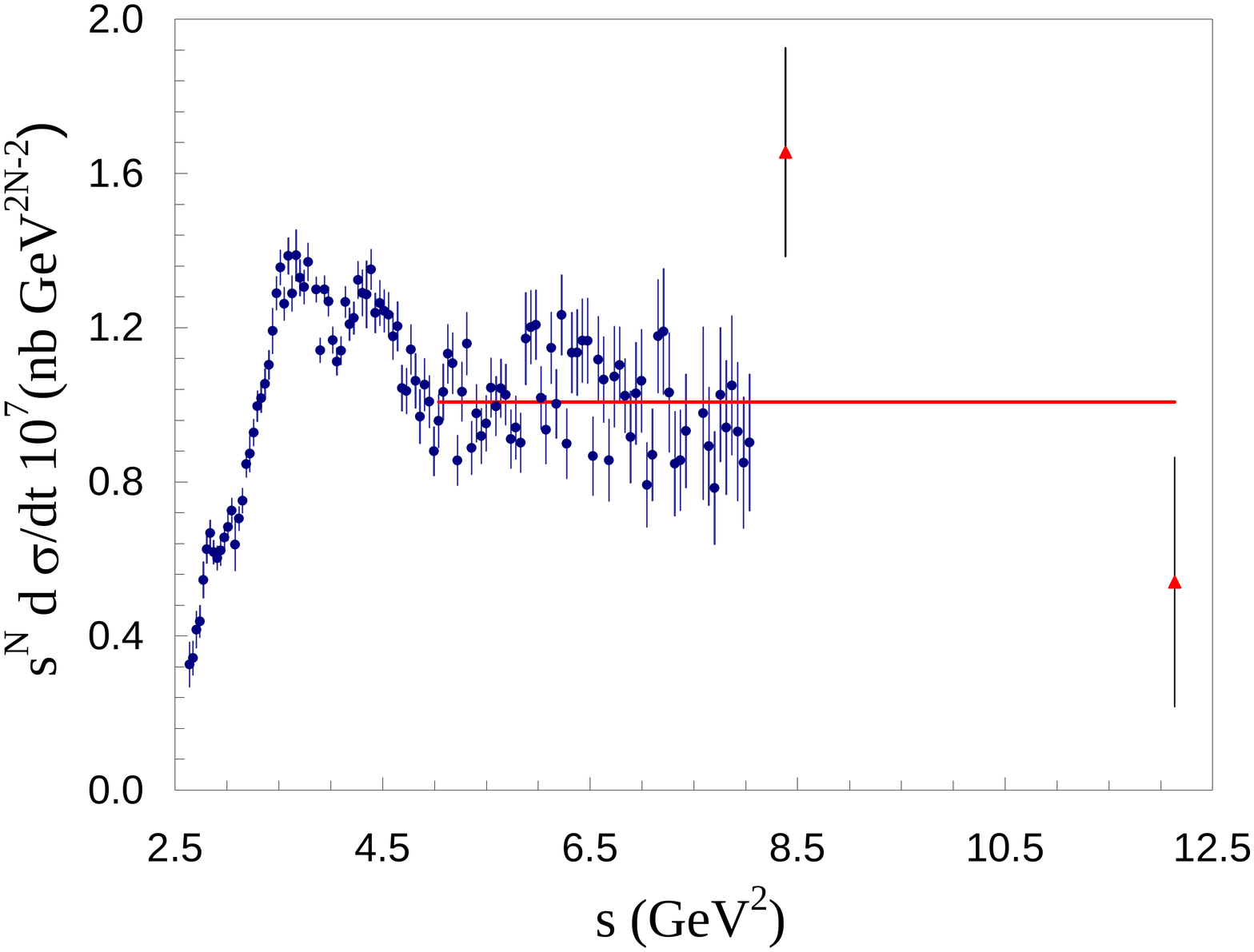}}
%\begin{figure*}
%\resizebox{1.0\textwidth}{!}{\includegraphics[angle=0.0]{Scaling_FigC.pdf}}
\caption{ (Color online) Cross section for the reaction $\gamma + p
\to K^+ + \Lambda$ at $90^\circ$ as a function of $s$ with a scaling
factor of $s^N$ applied.  CLAS data are solid blue
circles~\cite{McCracken:2009ra}, and SLAC data are red
triangles~\cite{Anderson:1976ph}.  The power-law fit (red line) is
discussed in the text.  }
\label{fig:dsdtN}       % Give a unique label
\end{figure}

The next question is to what degree the cross section satisfies the
constituent counting rule expectation of $s^{-7}$ scaling. The data
for $90^\circ$ is shown in Fig.~\ref{fig:dsdtN} scaled by a floating
power $N$ as $s^N$.  The choice of angle arises from where one expects
scaling to be apparent at the lowest $s$ while being furthest from
small $t$ and $u$.  A fit was performed to optimize $N$, and the value
of the scaled cross section is shown as the red horizontal bar.  For
$s > 5.0$ GeV$^2$ the CLAS data show a nearly flat behavior, while
below this there are bumps due to resonance production (to be
discussed in Sec.~\ref{sec:resonance}).  The best-fit values to
combined CLAS and SLAC data were found to be $N=7.1\pm0.1$ and
$s^N\frac{d\sigma}{dt} = 1.0 \pm 0.1$ nb GeV$^{2N-2}$, with
$\chi^2/\nu = 92/60$.  This is a fair-quality fit, and strongly
supports the validity of the $s^{-7}$ hypothesis that hinges on
counting the photon as a single bare elementary field.  In the
following discussion, we will take $N$ to be exactly 7.0.

In the range of $5<s<8$~GeV$^2$ where the scaling is observed, the
absolute cross section drops by a factor of 27, while the
$s^{7}$-scaled cross section varies between 0.8 to 1.2.  The onset of
scaling at $s\approx 5.3$~GeV$^2$ corresponds to produced mass
$W=2.3$~GeV, and we note that almost all the data points in the
present data set have transverse momentum in the center of mass
$p_\perp$ well below 1.0 GeV/c, averaging just 0.6 GeV/c.  All these
numbers indicate a much earlier onset of scaling as compared to the
$\gamma N\rightarrow \pi N$ channels~\cite{chen,zhu}, for which the
$s^{-7}$ scaling sets in at $W \ge 2.7$~GeV and $p_\perp \ge
1.2$~GeV/c.  This may indicate stronger convergence of the sum over
produced strange hadron states leading to the earlier onset of the
deep inelastic scattering regime relative to the case of non-strange
hadrons.  Possible small variation around the scaled value in
Fig.~\ref{fig:dsdtN} suggests the validity of local
duality\cite{zhaoclose,mek}; however, the study of the latter is
beyond the scope of this paper.

%%%%%%%%%%%%%%%%%%%%%%%%%%Fig 3%%%%%%%%%%%%%%%%%%%%%%%%%%%%%%%%%%%%
%remove the ``*'' to put the figures in-line with text.  
%\end{figure*}
%%%%%%%%%%%%%%%%%%%%%%%%%%Fig 3%%%%%%%%%%%%%%%%%%%%%%%%%%%%%%%%%%%%

\begin{figure}
\resizebox{0.50\textwidth}{!}{\includegraphics[angle=0.0]{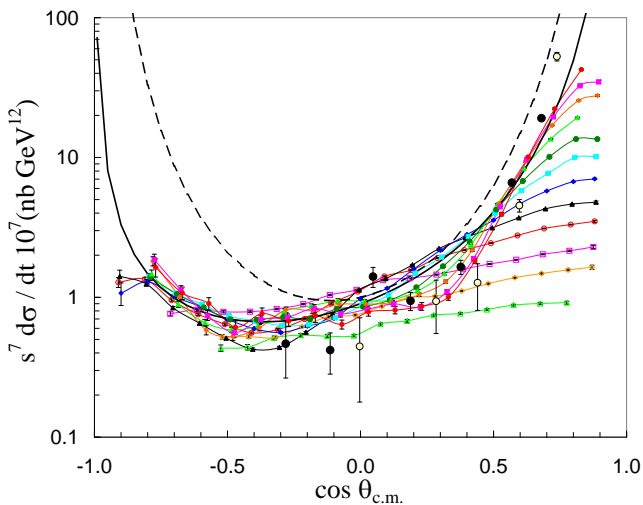}} 
%\begin{figure*}
%\resizebox{1.0\textwidth}{!}{\includegraphics[angle=0.0]{Scaling_FigH.pdf}}
\caption{ (Color online)
Scaled cross section versus center-of-mass meson production angle.
Each connected string of data points represents a 100 MeV wide band of
$W$, where the color and symbol coding is easiest to see on the
right-hand (forward angle) side.  From high to low: 
$W=2.78$ GeV (solid red circles),
2.68 GeV (solid magenta squares),
2.58 GeV (solid orange diamonds),
2.48 GeV (solid light green triangles),
2.38 GeV (solid dark green circles),
2.28 GeV (solid cyan squares),
2.18 GeV (solid blue diamonds),
2.08 GeV (solid black triangles),
1.98 GeV (open red circles),
1.88 GeV (open magenta squares),
1.78 GeV (open orange diamonds),
1.68 GeV (open light green triangles).
The CLAS data are the connected lines~\cite{McCracken:2009ra}, with
statistical errors that are usually smaller than the points.  The SLAC
data at $W=2.9$ GeV are are closed black circles, while 3.5 GeV data are open
white circles~\cite{Anderson:1976ph}. The curves are explained in the text.
}
\label{fig:ftheta}       % Give a unique label
\end{figure}

The evolution with energy of $f(t/s)$ in Eq.~\ref{eq:2} is shown in
Fig.~\ref{fig:ftheta}.  It presents the differential cross section
scaled by $s^7$ as a function of $\cos \theta_{c.m.}$.  In the high
energy limit, wherein masses can be ignored, one has $-t/s =
(1-\cos\theta_{c.m.})/2$.  Under the present kinematics this not the
case; for example at $90^\circ$ we have that $-t/s$ ranges from 0.12
to 0.36, not close to 0.5.  Nevertheless, we use $\cos
\theta_{c.m.}$ as a proxy for $-t/s$ to test for scaling.  Each
connected band of points shows the weighted mean of the data in a
range $\Delta W = 100$ MeV.  The bands range from near threshold,
centered at $W=1.68$ GeV, to a maximum $W = 2.78$ GeV.  It is evident
that the forward-angle scaled cross section rises rapidly with energy
but starts to plateau above about $W=2.6$ GeV.  For $\cos\theta_{c.m.} <
+0.5$, the bands converge when $W \ge 2.3$ GeV is reached. The
intermediate range of angles, which we shall take to be from $+0.1$ in
$\cos\theta_{c.m.}$ to $-0.5$, shows a fairly tight band of scaled
values at all $W$.  The small error bars show that the spacing between
the sets of points is very significant, and this energy dependence
will be discussed in Sec.~\ref{sec:resonance}.  The largest kaon
production angles show a uniform rise in the scaled cross section,
which we presume is evidence of $u$-channel contributions.  The rise
is less pronounced than what was observed in the $\pi N$ channels, but
similar to the $\pi \Delta$ channels~\cite{Anderson:1976ph}.  In
comparison to previous data from SLAC~\cite{Anderson:1976ph} at
$W=2.9$ GeV and 3.5 GeV, we see that the agreement is excellent; of
course the recent CLAS data have extended the precision and scope of
angle and energy coverage greatly.

The angular dependence is sensitive to the spin-isospin symmetry
structure of the valence quark wave function of the hadrons entering
in Eq.~\ref{eq:ampl} (see {\it e.g.} Refs.~\cite{KSPS,GS}).  In the case
of the photoproduction of pions, the amplitude probes the isospin
$\frac{1}{2}$ and $\frac{3}{2}$ and helicity $\frac{1}{2}$
combinations of the initial and final state partonic wave functions.
The $\gamma p \rightarrow \pi^+ n$ scattering yields rather symmetric
distribution around $\theta_{c.m.} =90 ^0$, consistent with the {\it
ad hoc} angular function~\cite{Anderson:1976ph}:
\begin{equation}
f^{\gamma p \rightarrow \pi^+n}(\cos \theta_{c.m.}) = 
(1-\cos\theta_{c.m.})^{-5}(1+\cos\theta_{c.m.})^{-4} 
\label{angle_pion}
\end{equation}
This is shown as the dashed line in Fig.~\ref{fig:ftheta}.  The
present data on $\gamma p \rightarrow K^+ \Lambda$ scattering, which
excludes isospin $\frac{3}{2}$ combinations, shows a markedly different
angular distribution consistent with
\begin{eqnarray}
f^{\gamma p \rightarrow K^+\Lambda}(\cos \theta_{c.m.}) = \hspace{4.5cm}
\nonumber \\ (0.9\pm.1)\times
(1-\cos\theta_{c.m.})^{-3.0\pm.2}(1+\cos\theta_{c.m.})^{-1.4\pm.1}
\label{angle_kaon}
\end{eqnarray}
on a scale of $10^7$ nb GeV$^{12}$.  This is shown as the solid black
line in Fig.~\ref{fig:ftheta}.  Evidently the $t$-channel (forward
angle) pieces of pion and kaon production are similar, while the
$u$-channel (back angle) portion of kaon production is less strong
than for pion production.  The $s^7$ scaling value of the cross
section averaged across {\it all} intermediate angles is
$(0.9\pm0.1)\times 10^7$ nb GeV$^{12}$.  This numerical value is the
same as the scaling value measured for $\gamma p\to \pi^+ n$ (see
Ref.~\cite{Anderson:1976ph,chen}).  This similarity of the scaled
cross section values was first noted by Anderson {\it et
al.}~\cite{Anderson:1976ph}.  This can be understood within the
framework of Eq.~(\ref{eq:ampl}), according to which if $|t|,|u| \gg
M_N^2,M_\Lambda^2$ such that all masses involved in the scattering can
be neglected, the hard kernel should have similar structure for
photoproduction of both pions and kaons.  Combining these results with
those for the $K^+\Sigma^0$ channel may allow, for example, to
constrain the relative weight of the vector and scalar diquarks in the
nucleon wave function (see {\it e.g.} Refs.~\cite{KSPS,GS}).  In
summary, one may say that the scaling function $f(t/s\to
\simeq\cos\theta_{c.m.})$ is approaching an energy-independent shape
for $W$ greater than $\approx 2.6$ GeV, but there remain significant
$\sim\pm30\%$ variations with energy at all angles.

%%%%%%%%%%%%%%%%%%%%%%%%%%Fig 4%%%%%%%%%%%%%%%%%%%%%%%%%%%%%%%%%%%%
%remove the ``*'' to put the figures in-line with text.  
%\end{figure*}
%%%%%%%%%%%%%%%%%%%%%%%%%%Fig 4%%%%%%%%%%%%%%%%%%%%%%%%%%%%%%%%%%%%

\begin{figure*}
\resizebox{1.0\textwidth}{!}{\includegraphics[angle=0.0]{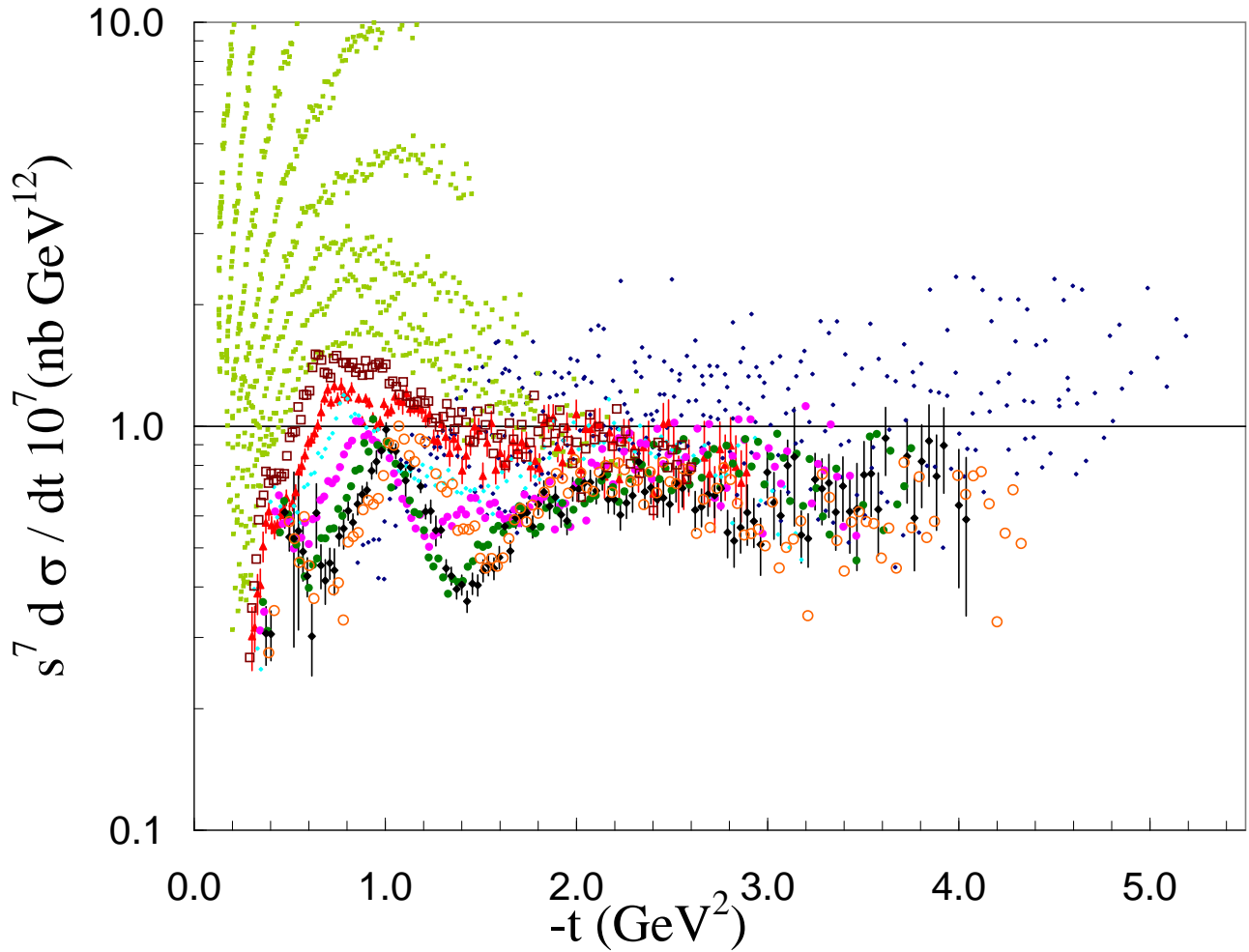}}
\caption{ (Color online)
Cross section for the reaction $\gamma + p \to K^+ + \Lambda$ as a function of
$-t$ with a scaling factor of $s^7$ applied.  Note how the mid- and
back- angle
points (not light green) fall approximately on a single locus.  The
data and color coding are the same as for Fig.\ref{fig:dsdt}.
}
\label{fig:dsdts7}       % Give a unique label
%\end{figure}
\end{figure*}

Figure~\ref{fig:dsdts7} again shows the whole data set scaled by
$s^7$, but versus $-t$.  The light green forward-angle data are now
``overscaled'', with clear positive-sloped bands corresponding, in
order from high to low values of $\cos\theta_{c.m.}$, of \{0.9,
0.8,...,0.2\}.  These forward angle bands were the ones that showed
the $N=2$ Regge scaling discussed above, so they clearly cannot also
exhibit $N=7$ scaling.  The intermediate angle data, however, do fall
on a roughly constant locus of points.  These intermediate angles are
color-coded for each angle bin as in Figs.~\ref{fig:dsdt} and
\ref{fig:dsdts2}, spanning the range $\cos\theta_{c.m.} = \{+0.1, 0.0,
-0.1, ... -0.5\}$ for all $W$.  Beside the overall $s^{-7}$ scaling of
the intermediate-angle cross sections, the other striking feature of
the scaled data in Fig.~\ref{fig:dsdts7} is its oscillatory
angle-dependent aspect between $\cos\theta_{c.m.}$ of $+0.1$ and
$-0.5$ for values of $-t$ from essentially zero to $-2.0$ GeV$^2$.
This behavior with angle seems strongly to suggest the interference of
resonant amplitudes.  Since the bump structure occurs not at fixed
$-t$, in Sec.~\ref{sec:resonance} we look instead at the structure as
a function of the invariant center of mass energy $W$.

%%%%%%%%%%%%%%%%%%%%%%%%%%Fig 5%%%%%%%%%%%%%%%%%%%%%%%%%%%%%%%%%%%%
%remove the ``*'' to put the figures in-line with text.  
%\begin{figure}
%\resizebox{0.50\textwidth}{!}{\includegraphics[angle=0.0]{Scaling_FigD.pdf}}
%%%%%%%%%%%%%%%%%%%%%%%%%%Fig 5%%%%%%%%%%%%%%%%%%%%%%%%%%%%%%%%%%%%

\section{\label{sec:resonance}RESONANCE CONTENT OF THE REACTION}

\begin{figure*}
\resizebox{1.0\textwidth}{!}{\includegraphics[angle=0.0]{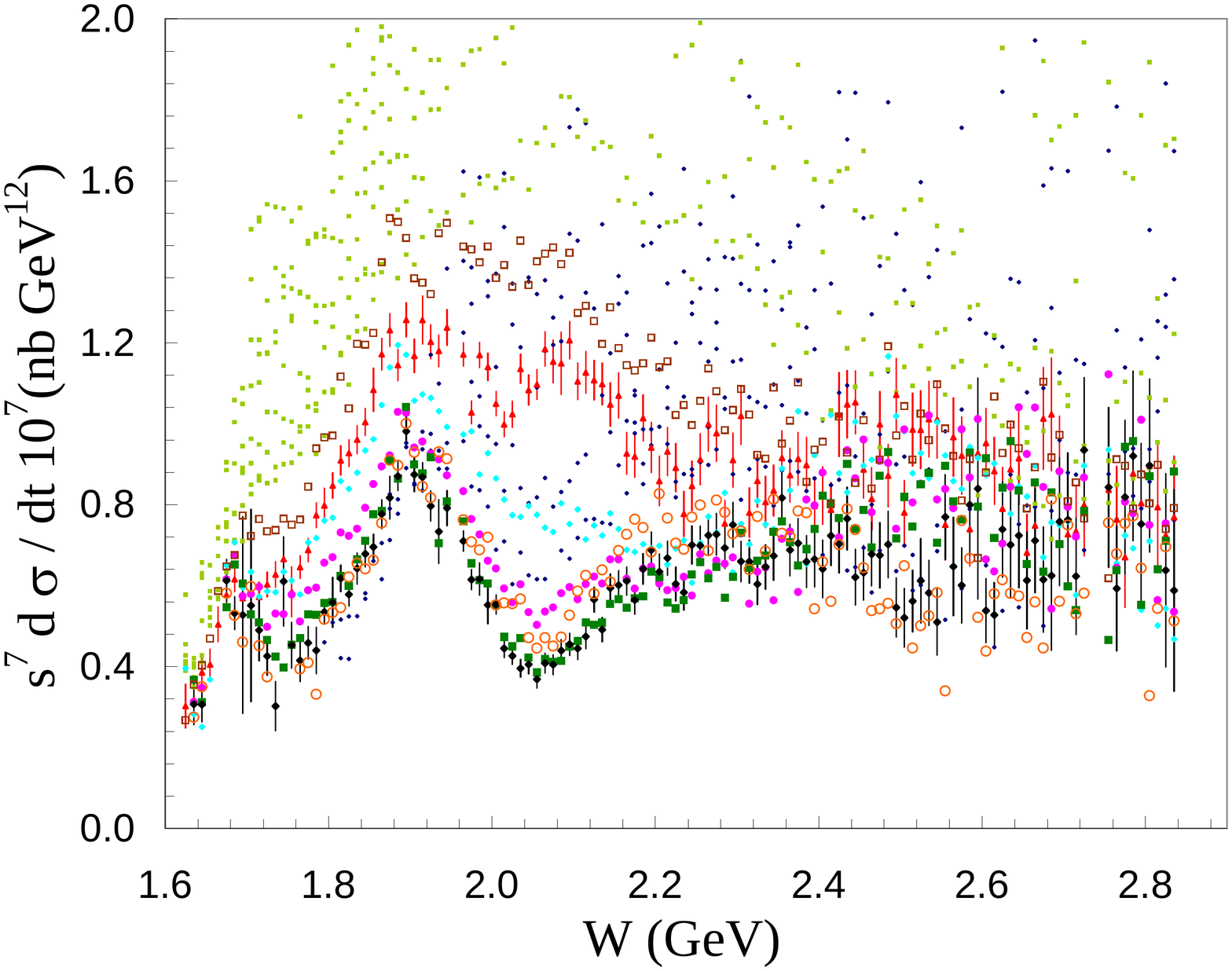}}
\caption{ (Color online) Cross section for the reaction $\gamma + p
\to K^+ + \Lambda$ as a function of $W$ with a scaling factor of $s^7$
applied.  Note how the mid- and back- angle bands of points (all but
light green) form distinctive features.  The vertical scale is now
linear rather than logarithmic.  The data and color-coding are the
same as for Fig.\ref{fig:dsdt}.  Representative statistical error bars
are shown for several angle bands only.  }
\label{fig:dsdts7w}       % Give a unique label
%\end{figure}
\end{figure*}

Figure~\ref{fig:dsdts7w} shows the scaled cross section again, but
plotted versus $W$ and on a linear scale.  The high-statistics
forward-angle (light green) points for $\cos\theta_{c.m.} \geq +0.2$
are now mapped across all $W$'s, and will henceforth be ignored, since
they clearly do not follow the $s^{-7}$ scaling trend. Also, we
similarly will ignore the low-statistics data at the largest
production angles (blue points) because of their low precision.  The
main observation here is that the oscillatory and interference-like
structures are now well-aligned near $W=$ 1.7, 1.9, and 2.1 GeV.  This
suggests that these structures are caused by $s-$channel resonance
production and interference.  The $s^7$ scaling has brought these
structures into sharp relief, even though they are clearly present
also in the unscaled data if one looks
carefully~\cite{McCracken:2009ra}.  The origin of the major peak at
$W=1.9$ GeV has been a source of debate for years, while the structure
near 2.1 GeV has gone unnoticed.  Resonance peaks at low $W$ were
observed previously in $s{^7}$-scaled pion
photoproduction~\cite{chen,zhu}, including an unexplained broad
enhancement near 2.1 GeV (See Ref.~\cite{chen}, Fig. 3). But the
present study is the first to discuss an angle-dependent interference
structure superimposed on the flat scaling plateau in hyperon
production.

%%%%%%%%%%%%%%%%%%%%%%%%%%Fig 6%%%%%%%%%%%%%%%%%%%%%%%%%%%%%%%%%%%%
%remove the ``*'' to put the figures in-line with text.  
%\begin{figure}
%\resizebox{0.50\textwidth}{!}{\includegraphics[angle=0.0]{Scaling_FigE.pdf}}
%%%%%%%%%%%%%%%%%%%%%%%%%%Fig 6%%%%%%%%%%%%%%%%%%%%%%%%%%%%%%%%%%%%

To investigate the nature of the resonance content seen in this
reaction, the following model of interfering resonance states was
developed to compare to the data.  Each $N^*$ resonance was modeled
with a relativistic Breit-Wigner amplitude written as
\begin{equation}
BW_{J_z}(m) = \frac{\sqrt{m  m_{0}\Gamma_{J_z,\gamma p \to N^*} \Gamma_{N^*\to
      K{\Lambda}}(q)}}{m^2 - m_{0}^2 - i m_{0}\Gamma_{tot}(q)},
\label{eq:bw}
\end{equation}
where $m=\sqrt{s}$ is the running mass value, $m_{0}$ is the centroid
mass of the resonance, and $q=q(m)$ is the c.m. 
frame momentum of the $K^+\Lambda$ final state.  The decay width to the
final state was written
\begin{equation}
\Gamma_{N^*\to K\Lambda} = \Gamma_{0}\left( \frac{q}{q_{0}}\right)^{2L+1}
\end{equation}
where $L$ is the orbital angular momentum of the decay, and $q_{0}$ is
the c.m. momentum at the resonance centroid energy.  The nominal decay
width $\Gamma_{0}$ was taken as an adjustable parameter.  The
photo-coupling to the $N^*$ state in spin projection $J_z$,
$\sqrt{\Gamma_{J_z,\gamma p \to N^*}}$, was taken as a complex
parameter.  The total width appearing in Eq.~\ref{eq:bw} was
\begin{equation}
\Gamma_{tot}(q) = \Gamma_{N^*\to K \Lambda}(q) + \Gamma_{S}(q),
\label{eq:gamtot}
\end{equation}
where $\Gamma_S(q)$ was designed to enforce the $s^{-7}$ scaling seen
in the data.  Without extra damping of the high-mass tails of the
Breit-Wigner line shapes, the computed cross section fits failed
utterly to reproduce the scaling.  Physically this may correspond to
the channel coupling and unitarity bounds that are ignored in this
model.  Therefore, an extra width, $\Gamma_S(q)$, was introduced in
the form:
\begin{equation}
\Gamma_{S} = \Gamma_{S_0}\left( \frac{q}{q_{S}}\right)^7.
\label{eq:gammahack}
\end{equation}
The reason for the power 7 in this expression is that at high energy
$q\to\frac{1}{2}\sqrt{s}$.  In the square of the matrix element,
therefore, the line shape scales asymptotically as $s^{-7}$.  The
parameter $\Gamma_{S_0}$ was fitted to be 0.50 GeV, which is the same
as the widths of the 4-star resonances in the 2 to 2.5 GeV mass
region.  The scale $q_{S}$ was a free parameter, chosen to make the
highest-$W$ portion of the curves have the correct behavior.  The
value turned out to be 0.77 GeV, which is larger that the values of
$q_0$ for any of the resonances included in the model.  Thus, this
phenomenological damping of the Breit-Wigner mass tails has an effect
at the high end of the mass distribution, and has little effect in the
region where the angle-dependent scaled cross section is prominent.

Using the beam axis as the quantization axis, each resonance was
allowed to couple to the unpolarized initial photon and proton states
via total spin projection $J_z = 1/2$ and $3/2$.  The final orbital
states that were allowed were $L=0,1,$ and $2$.  For example, the
final state amplitude $\psi_L(J,J_z)$ of a $J=3/2$ resonance formed
through the $J_z= 1/2$ initial spin projection and that decayed to a
$P$-wave final state was written as
\begin{equation}
\psi_P(\tfrac{3}{2},+\tfrac{1}{2})=\left\{\tfrac{1}{\sqrt{3}}Y_{1,1}\alpha_{\frac{1}{2},-\frac{1}{2}}+
%\nonumber \\
\sqrt{\tfrac{2}{3}}Y_{1,0}\alpha_{\frac{1}{2},+\frac{1}{2}}\right\} BW_{1/2}(m)
\end{equation}
where the $Y_{L,M}$'s are the spherical harmonics, and the 
$\alpha_{J,J_z}$'s are the nucleon spinors.  Analogous expressions
define the other final state amplitudes used:
$\psi_P(\frac{3}{2},+\frac{3}{2})$,
$\psi_D(\frac{3}{2},+\frac{1}{2})$,
$\psi_D(\frac{3}{2},+\frac{3}{2})$, and
$\psi_S(\frac{1}{2},+\frac{1}{2})$.  
The total angular intensity distribution as a function of $W
(=\sqrt{s}=m)$ and production angle, was then computed according to
\begin{eqnarray}
|A(m,\cos\theta_{c.m.})|^2 = |  
\psi_S(\tfrac{1}{2},+\tfrac{1}{2})+ \nonumber\\
\psi_P(\tfrac{3}{2},+\tfrac{1}{2})+
\psi_P(\tfrac{3}{2},+\tfrac{3}{2})+ \nonumber\\
\psi_D(\tfrac{3}{2},+\tfrac{1}{2})+
\psi_D(\tfrac{3}{2},+\tfrac{3}{2})|^2,
\end{eqnarray}
where we allow for a single $S$-wave, $P$-wave, and $D$-wave resonance only.

The unscaled cross section was computed using
\begin{equation}
\frac{d\sigma}{d\cos\theta_{c.m.}}(W,\cos\theta_{c.m.})=\frac{(\hbar
    c)^2}{32\pi}\frac{1}{s}\frac{q}{k} |A|^2,
\end{equation}
where the only additional factor is $k$, the initial state
center-of-mass momentum.  This cross section was then converted to
$d\sigma/dt$ and scaled by $s^7$, as before.

%%%%%%%%%%%%%%%%%%%%%%%%%%Fig 7%%%%%%%%%%%%%%%%%%%%%%%%%%%%%%%%%%%%
%remove the ``*'' to put the figures in-line with text.  
%\begin{figure}
%\resizebox{0.50\textwidth}{!}{\includegraphics[angle=0.0]{Scaling_FigF.pdf}}
\begin{figure*}
\resizebox{1.0\textwidth}{!}{\includegraphics[angle=0.0]{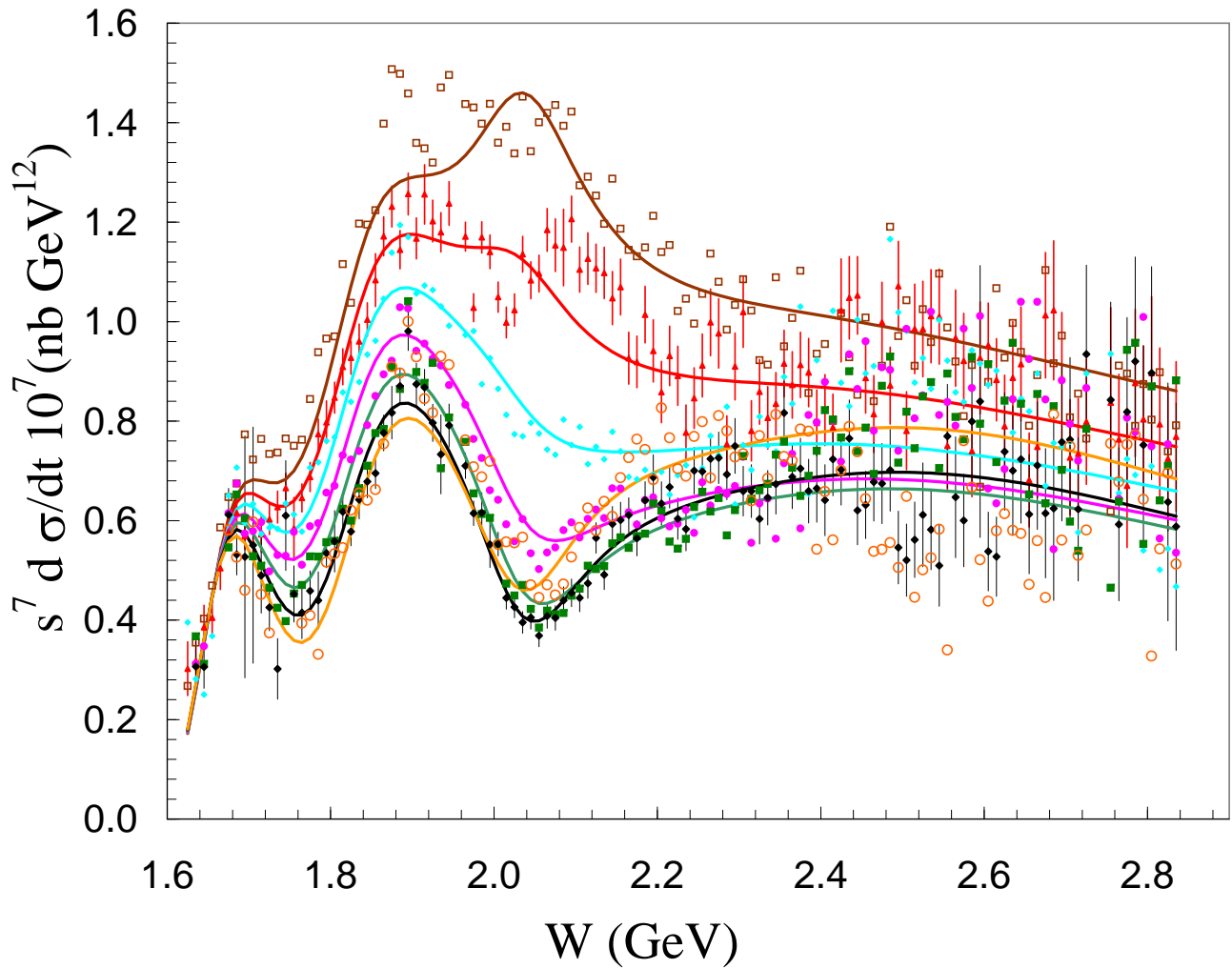}}
\caption{ 
(Color online) Cross section for the reaction $\gamma + p
\to K^+ + \Lambda$ as a function of $W$ with a scaling factor of $s^7$
applied, with model curves discussed in the text.
The color code is: $\cos\theta_{c.m.}$ = $+0.1$ (open brown squares), $0.0$
(filled red triangles), $-0.1$ (filled cyan points), $-0.2$ (filled
magenta circles), $-0.3$ (filled green squares), $-0.4$ (filled black
diamonds), $-0.5$ (open orange circles).  Representative statistical
error bars are shown for several angle bands only.  
}
\label{fig:dsdtfit}       % Give a unique label
%\end{figure}
\end{figure*}
%%%%%%%%%%%%%%%%%%%%%%%%%%Fig 7%%%%%%%%%%%%%%%%%%%%%%%%%%%%%%%%%%%%

Various combinations of total $J$ and decay waves were tested to find
a reasonable representation of the data.  The most successful
combination is shown in Fig.~\ref{fig:dsdtfit}.  The large and small
angle data points have now been suppressed, as justified above, and
the intermediate-angle data are shown together with the corresponding
line shapes from the parametrization given above.  The best
combination of waves found was an $S_{11}$ resonance near threshold, a
$P_{13}$ resonance centered at 1.92 GeV, and a $D_{13}$ resonance
centered at 2.10 GeV.  Other combinations of waves with various values
of $L$ and $J$ were tested, but each of those resulted in unacceptable
angular distributions.  For the high-mass $D_{13}$ state at 2.10 GeV,
alternatives tested were $P_{13}$, $P_{11}$, $S_{11}$ and $D_{15}$.
The final tabulated values for the centroids, width, and couplings are
given in Table~\ref{resulttable}, together with an estimate of the
uncertainties.  The photo-couplings are given as the magnitude in
(GeV)$^{1/2}$ and phase in degrees, as specified in Eq.~\ref{eq:bw}.
These values are the result of exploratory fits for a single reaction
channel, with possible finer details in the data not reproduced.  For
instance, we did not include the $N(1720)P_{13}$ that is
established~\cite{Anisovich:2010an} in multi-channel fits, but which
is not dominant near threshold.  The identification of the principal
resonant components in the fits seems secure, however.  In particular,
taking the state at 1.9 GeV to be $P_{13}$ leads to the state at 2.1
GeV being strongly favored as $D_{13}$.

\begin{figure}
\resizebox{0.50\textwidth}{!}{\includegraphics[angle=0.0]{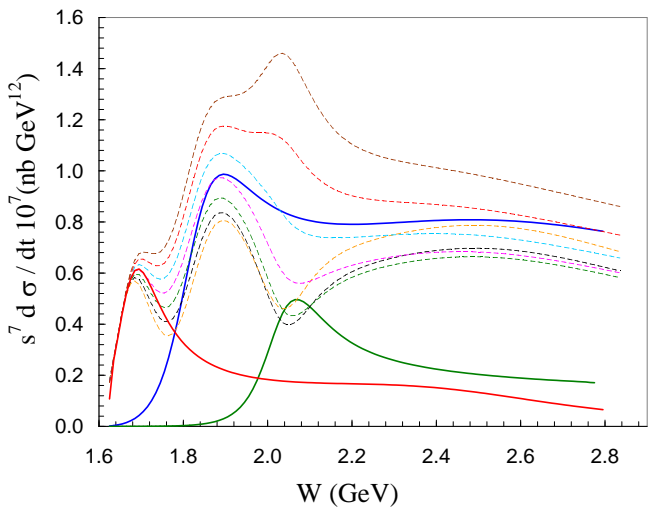}}
%\begin{figure*}
%\resizebox{1.0\textwidth}{!}{\includegraphics[angle=0.0]{Scaling_FigG.pdf}}
\caption{ (Color online) Phenomenological model for the reaction
$\gamma + p \to K^+ + \Lambda$ as a function of $W$ with a scaling
factor of $s^7$ applied.  The upper dashed curves are for individual
angles as in Fig.\ref{fig:dsdtfit}, while the lower curves show the
components: $S$-wave (solid red), $P$-wave (solid blue), $D$-wave
(solid green).  }
\label{fig:dsdtshape}       % Give a unique label
\end{figure}

Figure~\ref{fig:dsdtshape} shows the final line shapes again, but with
the underlying resonance shapes included.  Note the long tails on each
modified Breit-Wigner curve that arises from the interplay of the
scaling by $s^7$ and the damping width specified in
Eq.~\ref{eq:gammahack}.

%%%%%%%%%%%%%%%%%%%%%%%%%%Fig 8%%%%%%%%%%%%%%%%%%%%%%%%%%%%%%%%%%%%
%remove the ``*'' to put the figures in-line with text.  
%\end{figure*}
%%%%%%%%%%%%%%%%%%%%%%%%%%Fig 8%%%%%%%%%%%%%%%%%%%%%%%%%%%%%%%%%%%%

\begin{table}
\caption{ Results for the resonant content fitted to the scaled cross
section.  The masses and widths are for the fitted relativistic
Breit-Wigner functions.  The overall phases are specified relative to
the $P_{13}$ state.  }
\begin{center}
\begin{tabular}{|c|c|c|c|c|}
\hline
\hline
Resonance & $m_0$  & $\Gamma_0 $ & $\sqrt{\Gamma_{1/2,\gamma p\rightarrow N^*}}$ & $\sqrt{\Gamma_{3/2,\gamma p\rightarrow    N^*}}$ \\ 
\&Decay   & (GeV)  &   (MeV)     & (GeV)$^{1/2}$                                 & (GeV)$^{1/2}$ \\ 
          &        &             & Phase                                         & Phase \\ 
\hline 
$S_{11}$ & $1690 \pm 10$ & $ 80\pm 20$ & $1.83 \pm.10$         &                    \\
         &               &             & $(-142 \pm 5)^\circ$  &                    \\
$P_{13}$ & $1920 \pm 10$ & $440\pm100$ & $1.93 \pm.10$         & $1.67 \pm.07 $     \\
         &               &             &       $-$             &      $-$           \\
$D_{13}$ & $2100 \pm 20$ & $200\pm 50$ & $0.61 \pm.10$         & $1.19 \pm.10 $     \\
         &               &             & $ (45 \pm 5)^\circ$   & $ (45 \pm 5)^\circ$\\

\hline
\hline
\end{tabular}
\end{center}
\label{resulttable}
\end{table}

\section{\label{sec:conclusion}DISCUSSION AND CONCLUSIONS}

We have discussed three features of the $\gamma + p \to K^+ + \Lambda$
reaction.  First, we have confirmed~\cite{bradforddsdo}, with higher
statistics~\cite{McCracken:2009ra}, the Regge-domain scaling with the
power $s^{-2}$ for the low-$t$ data.  Second, we have shown that the
constituent counting rule prediction of $s^{-7}$ scaling is quite well
satisfied for this reaction for $|-t| \ge 2$ GeV$^2$, for $W \ge
2.3$~GeV, and $p_\perp \ge 0.8$~GeV/c.  These values are much lower
than those observed for $\pi N$ photoproduction.  The scaled cross
sections for $\pi^+ n$ and $K^+ \Lambda$ are equal at $90^\circ$ , but
the angular dependencies away from $90^\circ$ differ
(Fig.~\ref{fig:ftheta}). Quantitative analysis of this scaling based
on QCD models or an alternative description based on unitary channel
coupling remains to be investigated.

Third, we have shown that the scaled cross section at low to moderate
$-t$ and $W < 2.3$ GeV shows structure consistent with the
interference of a set of $s$-channel resonances.  We make the ansatz
that the scaled cross section can be analyzed with interfering
Breit-Wigner resonance line shapes with the introduction of an
empirical damping factor to achieve the $s^{-7}$ scaling limit; the
qualitative appearance of the data suggests this is reasonable.
With this assumption, the best match to the data was found to include
an $N(1690)S_{11}$, an $N(1920)P_{13}$ and an $N(2100)D_{13}$ suite of
resonances.  We remark that previous recent work by several groups
using various effective Lagrangian formulations have consistently
required an $S$-wave near threshold and either a $P$- or $D$-wave near
1.9 GeV.  The present work shows that the structure at 1.9 GeV is
consistent with a $P_{13}$ state.  No previous work has made a claim
for a $D$-wave structure near 2.1 GeV in this reaction channel;
previously, data were too sparse to clearly examine the angular
dependence in this mass range.  From Fig.~\ref{fig:dsdtshape} one sees
that this state influences the reaction also below 2 GeV.  Hence,
earlier work that concentrated on $W$ below 2 GeV may need revision.
We also note that the ``2 star''~\cite{Nakamura:2010zzi} resonance
$N(2080)D_{13}$ was identified by Capstick and
Roberts~\cite{Capstick:1998uh} in their relativized quark model as
having large combined strength in both photo-coupling and decay to
$K\Lambda$.  This led Mart and Bennhold~\cite{mart} to tentatively
identify this state with the large structure at 1.9 GeV.  Our study
supports the existence of a $D_{13}$ state coupling to $K^+\Lambda$,
not at 1.9 GeV but rather at 2.1 GeV.

In this paper we have ignored the available spin observables for this
reaction, including the recoil polarization of the $\Lambda$ hyperon,
$P_\Lambda$ ~\cite{McNabb:2003nf,McCracken:2009ra}, the beam asymmetry
$\Sigma$ ~\cite{leps_zegers}, the beam-recoil double polarizations
$C_x$ and $C_z$ ~\cite{Bradford:2006ba,Schumacher:2008xw}, and the
beam-recoil linear double polarizations $O_x$ and $O_z$
~\cite{Lleres:2007tx}.  Including these in amplitude-level fits
results in much more sensitivity to smaller contributions to the
reaction mechanism.  Further, we have ignored the effects of unitarity
bounds and channel coupling.  Thus, the main result of this work has
been to demonstrate how two different types of scaling apply to this
reaction, and to demonstrate how the $s^7$-scaled cross section
highlights some of the important resonance contributions to the
reaction mechanism.

This phenomenological analysis of the $\gamma + p \to K^+ + \Lambda$
reaction has thus yielded some insights into this reaction.  However,
a theoretical foundation for using the notions of $s^{-7}$ scaling and
baryon resonance analysis in the same framework is lacking.  With the
observations made here, we hope to stimulate further efforts to
understand whether this approach can be put on a more rigorous
footing.  We expect to study other meson photoproduction reactions to
test the consistency of this approach.

\begin{acknowledgments}
This work was supported by US Department of Energy grants
DE-FG02-87ER40315 and
DE-FG02-01ER41172.
\end{acknowledgments}
\vfill

%-------------------------------------------------------

\end{document}